\documentclass{sigchi}

\toappear{}

\usepackage{balance}  
\usepackage{graphicx} 
\usepackage{times}    
\usepackage{url}      
\usepackage{lscape}  
\usepackage{epstopdf}
\usepackage{float}
\usepackage{color}
\usepackage{amsmath}  
\usepackage{mathptmx}
\usepackage{listings}
\usepackage{lscape}  
\usepackage{nicefrac}
\usepackage{textcomp}
\usepackage{fixltx2e}

\makeatletter
\def\url@leostyle{%
  \@ifundefined{selectfont}{\def\UrlFont{\sf}}{\def\UrlFont{\small\bf\ttfamily}}}
\makeatother
\def\pprw{8.5in}
\def\pprh{11in}

\usepackage[pdftex]{hyperref}
\hypersetup{
pdftitle={SIGCHI Conference Proceedings Format},
pdfauthor={LaTeX},
pdfkeywords={SIGCHI, proceedings, archival format},
bookmarksnumbered,
pdfstartview={FitH},
colorlinks,
citecolor=black,
filecolor=black,
linkcolor=black,
urlcolor=black,
breaklinks=true,
}

\begin{document}

\title{Where's My Drink? Enabling Peripheral Real World Interactions While Using HMDs}
\vspace{-10pt}

\numberofauthors{1}
\author{
  \alignauthor Pulkit Budhiraja, Rajinder Sodhi, Brett Jones, Kevin Karsch, Brian Bailey, David Forsyth\\
    \affaddr{University of Illinois at Urbana-Champaign, Urbana, Illinois, United States}\\
    \email{\{budhirj2, rsodhi2, brjones2, karsch1, bpbailey, daf\} @illinois.edu }\\
}
\vspace{-10pt}

\maketitle

\begin{abstract}
Head Mounted Displays (HMDs) allow users to experience virtual reality with a great level of immersion. However, even simple physical tasks like drinking a beverage can be difficult and awkward while in a virtual reality experience. We explore mixed reality renderings that selectively incorporate the physical world into the virtual world for interactions with physical objects. We conducted a user study comparing four rendering techniques that balances immersion in a virtual world with ease of interaction with the physical world. Finally, we discuss the pros and cons of each approach, suggesting guidelines for future rendering techniques that bring physical objects into virtual reality.
\end{abstract}


\category{H.5.1.}{Information Interfaces and Presentation (e.g. HCI)}{Artificial, augmented, and virtual realities}

\section{Introduction}

HMDs are now widely available so that consumers can enjoy a variety of Virtual Reality (VR) experiences in their living rooms. While being highly immersive, HMDs occlude the real world making physical and social interactions difficult and awkward. Currently, users have two choices: keep the HMD on and blindly interact with the world, or take the HMD off and break their immersion to the virtual experience. Such context switching between worlds is expensive: it takes time to be immersed in a virtual environment ~\cite{Brown2004}, and frequent switching between worlds can be disorienting. 

\begin{figure}[h]
	\centering
	\includegraphics[width=3.35in]{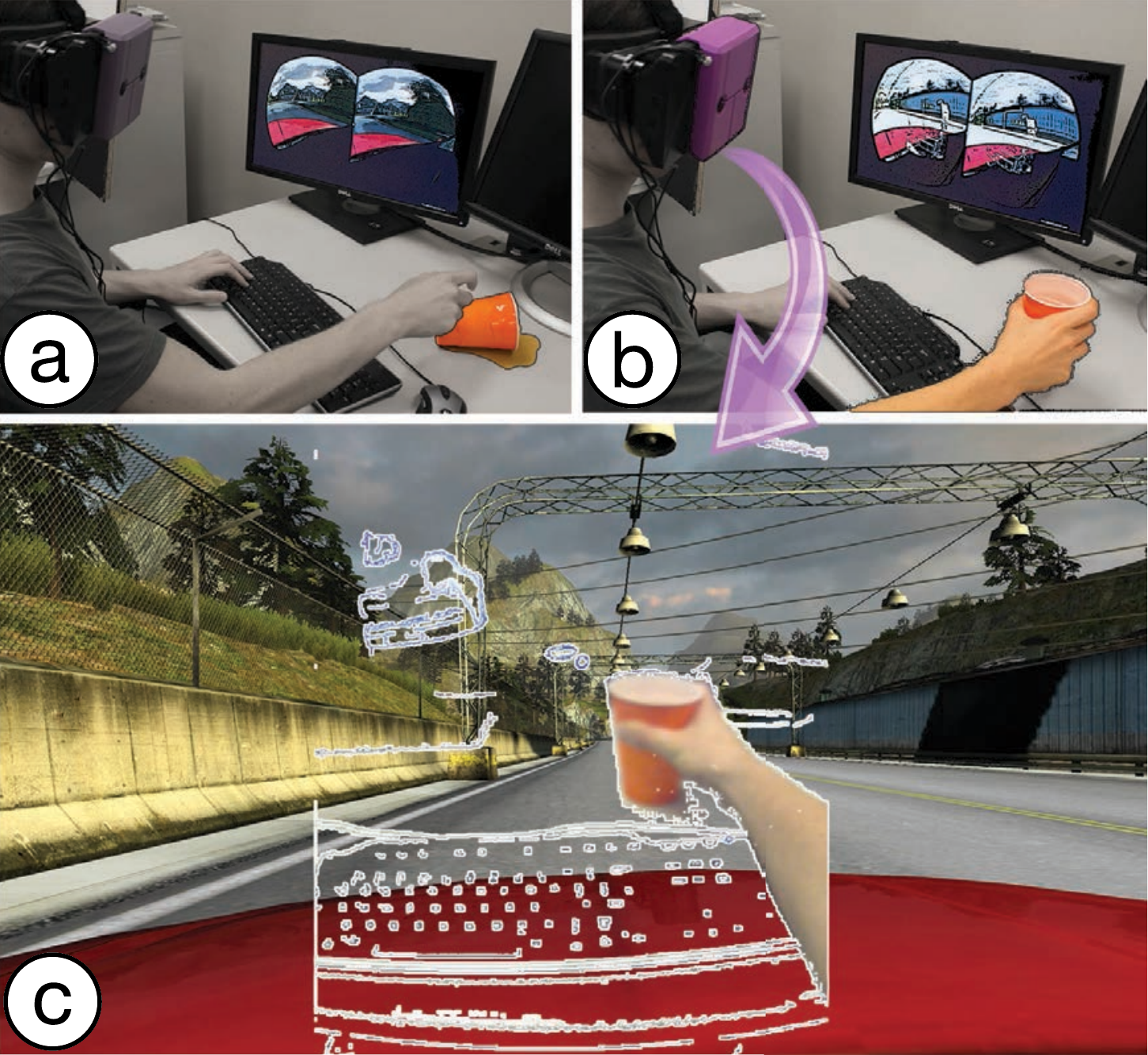}
	\caption{ Performing real world, peripheral tasks while using VR HMDs can be (a) frustrating and messy. Our system, (b) comprised of two inexpensive webcams and (c) augmented virtuality renderings, allows users to perform peripheral tasks, such as grabbing a drink, while still being immersed in the virtual experience.\vspace{-8pt}}
	\label{fig:teaser}
\end{figure}
\begin{figure*}[tb]

	\centering
	\includegraphics[width=7in]{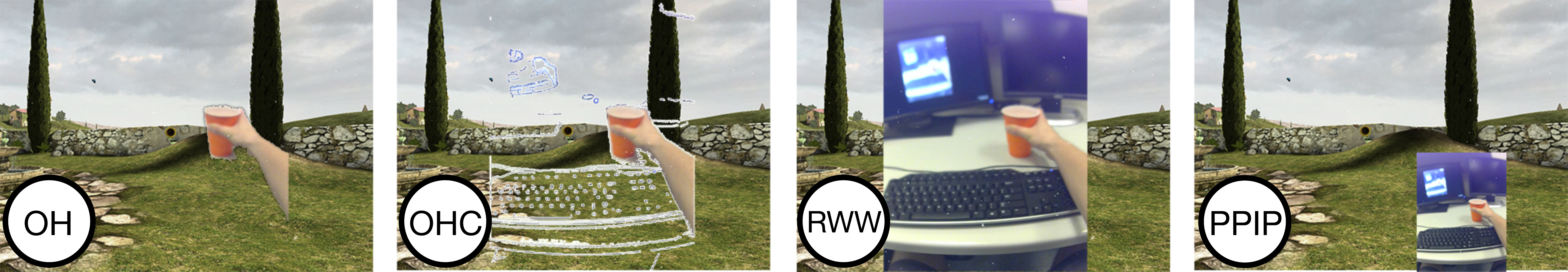}
	\caption{ All the different renderings. (a) Object \& Hand, (b) Object, Hand \& Edges, (c) Real World Windowed, (d) Physical Picture In Picture}
	\label{fig:renderings}

\end{figure*}
In this paper, we explore a design space of rendering techniques that enable users wearing an HMD to interact with the physical environment. Our goal is to make interactions with the physical world more seamless, while keeping the user immersed in the virtual world. Unlike previous work in augmented reality \cite{Bai2014, Gordon} that explores using stereo cameras to superimpose virtual objects in the physical world, we overlay physical objects on top of a virtual environment (i.e. augmented virtuality~\cite{Milgram1994}). Users can see their hands in the virtual environment, peripheral objects (e.g. a cup), or even perform social interactions with co-located players (see Figure~\ref{fig:teaser}). 


We evaluate this design space of mixed reality renderings with a user study comparing different renderings of varying visual fidelity across different virtual experiences (a movie, a first person shooter, and a racing game). Our results show that users prefer renderings which selectively blend virtual and physical, while maintaining a one-to-one scaling of the physical environment. The highest rated rendering allows users to see their hands, objects of interest and salient edges of the surrounding environment.

This paper's contributions are two fold: (1) we explore a series of rendering techniques that uses a stereo camera pair to selectively incorporate aspects of the physical environment into the virtual experience, and (2) we show results from a user study comparing the renderings and discuss the benefits and limitations of each approach.

\section{Related Work}
Since its inception, a large body of VR work has explored incorporating virtual models of physical objects into virtual experiences. For instance, a low-fidelity model of a user's hands could be captured using data input gloves \cite{Mine1997}. Previous work has also explored approaches for directly merging the virtual world with the physical world~\cite{Milgram1994}. Augmented Reality (AR) overlays virtual objects into the physical world, and has a rich history of use in mobile phones and HMDs (e.g., \cite{Billinghurst2002}). In contrast, our work is aligned with Augmented Virtuality (AV), where virtual reality is enhanced with parts of the physical world, grounding the experience in the virtual world. Previous work in AV has focused on collaborative applications including displaying real world video on virtual office windows~\cite{Karl-petter1997} or displaying group communication around a virtual table~\cite{Regenbrech2004}. More recent work has explored physical depth based renderings of a user's hands in VR for productivity applications ~\cite{Kreylos2014}. In contrast, we focus on peripheral physical interactions, exploring a design space of rendering techniques that selectively show aspects of the physical world, reinforcing immersion while minimizing distraction.

\section{Design Space}
We highlight a design space that focuses on tradeoffs between awareness of the physical world while remaining focused on game play (see Figure~\ref{fig:renderings}). We selectively identify and render aspects of the physical world that provide users with varying amounts of information of the physical space. These renderings are not exhaustive and gave us an initial starting point for exploring this design space. The renderings are best understood by demonstrations (please see the accompanying videos).

\subsection{Renderings}
\emph{\textit{Object \& Hands (OH)}} : The first rendering shows only the object of interest and the user’s
hands. This is the minimal information necessary to maintain
proprioceptive feedback \cite{Mine1997}. This rendering enables the
user to focus on the virtual experience, at the expense of limited
knowledge of the physical environment.

\emph{\textit{Object, Hands \& Context (OHC)} }: The second rendering shows the object of interest, the user’s hands
and surrounding physical objects with edges. This rendering provides
additional context at the expense of potential distraction from the
virtual experience.

\emph{\textit {Real World Windowed (RWW)} }: The third rendering provides a windowed view of the physical world,
with the virtual world still shown in the user's peripheral vision.
The real world is rendered in a fully opaque window at the center of
the user's visual field. This rendering allows the user to focus on
their interactions in the physical world, while still maintaining
peripheral cues about the virtual environment.

\emph{\textit {Physical Picture in Picture (PPIP)}} : The fourth rendering shows the physical world as a picture in picture
rendering in the lower right hand corner of the screen (small version
of the virtual world), mimicking the behaviour of picture-in-picture
televisions. This rendering allows users to interact with the physical
world, without taking up as much screen real-estate as \emph{RWW}.

\section{User Study}

The purpose of this study was to elicit qualitative feedback about the design space of renderings in the context of different genres of VR experiences. We specifically wanted to evaluate if our renderings allow users to remain immersed in the virtual experience while seeing parts of their physical environment. We also compared our renderings to the status quo (baseline) solution for interacting with the physical environment while wearing an HMD, namely to remove the HMD entirely. 

Given the wide variety of VR experiences, we evaluated a spectrum of experiences that vary in the level of user engagement. Some VR experiences are entirely passive and require no input from the user (watching a movie), and other experiences require continuous attention and high levels of user input (a racing game). We hypothesize that (1) the preferred rendering will depend on the virtual content, changing with varying levels of user engagement, and (2) that users will prefer OH since it provides a balance between visual information of the physical world, without being overly distracting. 

\begin{figure*}[t]

	\centering
	\includegraphics[width=7in]{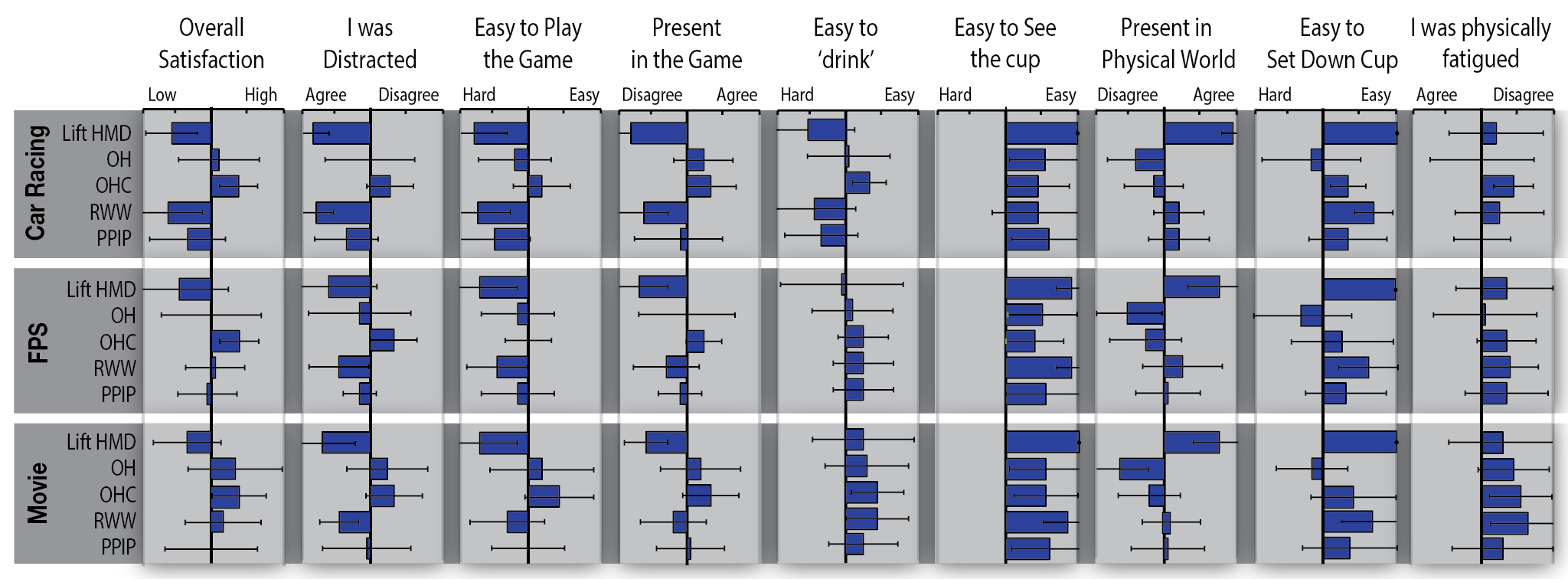}
	\caption{ Mean ratings for each method evaluated in the user study. Participants rated 'overall satisfaction' and 8 other factors on a 5 point likert scale. The errorbars represent standard deviations. OHC is highly rated for overall satisfaction in all 3 VR environments ($\mu$\textsubscript{OHC-Movie} = 3.7,  $\mu$\textsubscript{OHC-FPS} = 3.7, $\mu$\textsubscript{OHC-Car} = 3.7). }
	\label{fig:interTaskGraph}

\end{figure*}

\subsection{Virtual Scenarios}
We selected rich visual experiences that are representative of real-world use cases in VR. We created the following experiences in Unity3D: (1) watching a movie~\url{http://sintel.org} in a movie theater, (2) a First Person Shooter (FPS) modified from Unity3D's 3rd person AngryBots sample, and (3) a racing game modified from Unity3D's Car Tutorial.


The movie is a passive experience with no user input that uses a limited field of view and requires minimal head movement. The FPS is fast paced and requires both mouse and keyboard input with lots of head motion, but still contains natural pauses in game play for the user to interact with the physical environment. The racing game is a continuous attention task, where the user must constantly steer their car or risk crashing, with only keyboard input needed leaving one hand free to interact with the physical environment.


\subsection{Physical Setup}

Our system comprises of an HMD with stereo cameras and vision processing. We use the Oculus Rift DK1, augmented with 2 Logitech C310 webcams to provide a stereoscopic view of the real world. The lenses of the cameras were replaced with 1.8mm lenses (\url{http://www.thingiverse.com/thing:305355}) to provide a wider FOV of approximately 120 degrees, and then mounted in a 3D printed mount (\url{http://www.thingiverse.com/thing:323913}). Objects and hands were segmented using color based segmentation. Users also wore headphones to create a fully immersive experience.


\subsection{Participants}
We recruited 16 subjects (13 male), ages 18-24 years with some PC gaming experience and corrected to normal vision. Of our 16 participants, 6 were excluded from the study due to signification simulator sickness (even with brief pauses between conditions). Simulator sickness is common with current HMDs~\cite{Kennedy2010494}, and only 10 participants completed the study, producing a total of 300 grasping trials. We expect that improvements in display latency, resolution and refresh rate will decrease simulator sickness in the near future.

\subsection{Tasks and Procedure}
We designed a within subjects user study, where subjects interacted with 3 VR experiences in a randomized ordering using a PC keyboard and mouse while seated at a table. While engaged in the virtual experience, subjects were externally prompted every 45-60 seconds to pick up a physical cup of water, drink the water (or simulate), and place the cup back on the table. This physical task was repeated twice for each of the 4 rendering methods and the baseline method, in a randomized ordering with 5 minutes of rest between renderings. Subjects were instructed to focus on their performance in the VR experience, simulating real-world conditions where users are highly engaged in the game/movie. Physical distractor objects were included on the table as well (a mobile phone, speakers and pieces of paper). After each physical interaction with the cup, the experimenter moved the cup to simulate the user loosing track of the physical environment during more realistic long term play scenarios. In total there were 30 trials per subject, 3 VR experiences x 2 repetitions x (4 renderings + baseline).

Between each rendering, subjects completed a questionnaire inspired by the core modules of~\cite{Schild2012}, where they rated their overall satisfaction, immersion, level of distraction, ease of play etc. (see Figure \ref{fig:interTaskGraph}). At the end of the study, subjects ranked the rendering methods along various dimensions (see Figure~\ref{fig:meanRanks}), with visual mnemonics to remind the users of each condition. Finally, we conducted a semi-structured interview with think-aloud subject feedback.

\subsection{Results}
The intra-rendering results (see Figure \ref{fig:interTaskGraph}) show an overwhelming support for OHC, which was rated as the most preferred method by participants across all VR scenarios ($\mu$\textsubscript{OHC-Movie} = 3.7,  $\mu$\textsubscript{OHC-FPS} = 3.7, $\mu$\textsubscript{OHC-Car} = 3.7). A Kruskal Wallis non-parametric test found significant differences between visual renderings. A post-hoc Bonferroni-corrected Wilcoxon test on the OHC performed significantly better than RWW and PPIP, both in Car (Z = -2.713, p \textless 0.01, Z = -2.56, p \textless 0.01) and FPS (Z = -1.732, p \textless 0.01, Z = -1.99, p \textless 0.01). 

\begin{figure}[tb]

	\centering
	\includegraphics[width=3.35in]{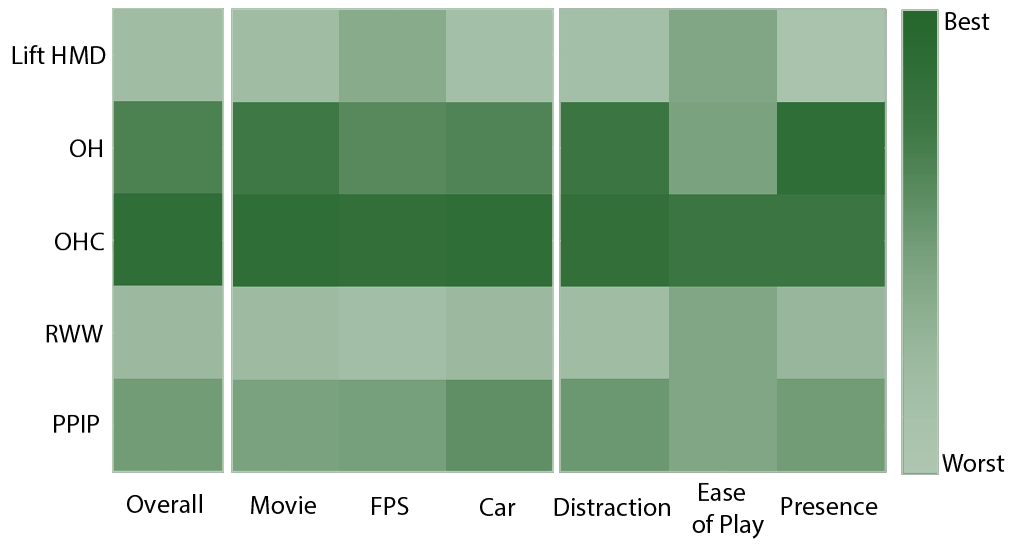}
	\caption{ Post user study mean rankings. (left to right) Overall satisfaction across all VR experiences, overall satisfaction for each VR experience (Movie, FPS, Car), overall distraction, ease of play and presence across all VR experiences.\vspace{-8pt} } 
	\label{fig:meanRanks}

\end{figure}

This result was further validated in the mean rankings analysis where participants consistently ranked OHC highest in overall satisfaction across all VR scenario and also for each individual VR scenario (Figure \ref{fig:meanRanks}). The baseline condition of removing the HMD was always the least preferred approach. However, Figure \ref{fig:interTaskGraph} illustrates a substantial pattern where Lift HMD, RWW and PPIP were more acceptable to participants in Movie, eventually becoming less tolerable in the higher engagement scenarios (FPS \& Car). Contrary to our expectation of OH being ranked the best method, OH was consistently ranked second in overall satisfaction, immersion, presence and distraction with high variance across users ($\sigma$\textsubscript{OH-Movie} = 1.174,$\sigma$\textsubscript{OH-FPS} = 1.247, $\sigma$\textsubscript{OH-Car}  = 1.033).

On the whole, participant’s qualitative feedback reflected our empirical findings. With OHC, users described how the additional contextual information of their surroundings aided them in finding the cup. As one user described, \emph{``The extra lines helped me find the cup and put it back, and it wasn’t distracting, it didn’t break my concentration on the game.''} However, some participants liked the minimal nature of OH, with one user commenting \emph{``even though I could not see the cup...I got used to the surroundings and the table and had a fair idea of where to look for the cup.''} In contrast, one user who favored OHC noted for OH,\emph{``I felt lost and had to feel the physical space around me to look for the cup.''}

The Lift HMD approach was disliked across VR scenarios with one participant commenting, \emph{``removing the goggles is immersion breaking, with [OH] and [OHC], I still felt pretty much in the game. [RWW] and [PPIP] are a little more immersion breaking.''} With RWW, one user commented, \emph{``you are kind of in limbo when you’re doing [RWW], you might as well lift and do it quickly. I don’t feel like I am part of the virtual world but I don’t feel like I am in the real world.''} When asked if participants would prefer any other location for rendering the preview window in PPIP, one user noted, \emph{``it wouldn’t make any significant difference since you still have to concentrate on a corner which takes away your focus from the game.''}

\section{Discussion \& Future Work}
The clear ‘winner’ among our selection of visual renderings was to show users the object of interests and their hands while using edges to visualize the supporting surfaces. OHC allowed users to quickly re-acclimate themselves to the physical environment, particularly when significant head/body motions disconnected users from their physical surroundings. Some users suggested that pausing the game would be preferred, however this is only possible with non-multiplayer games. Future work could explore various methods for pausing the game experience via audio input, touch input on the HMD, controller input or even automatically detecting a user's reaching motion.

Designers looking to visualize aspects of the physical world should consider balancing the scale of the rendered objects with its placement in the virtual world. We found that users felt naturally comfortable seeing a version of the cup that was close in size to the actual physical cup and from their own ego-centric viewpoint. This was not the case with the PPIP technique which forced users to switch between seeing the game and the window while requiring additional time to acclimate to the small sized view of the physical world. Furthermore, visual rendering techniques could be designed in the future to take advantage of unused pixels in the virtual environment. For example, users frequently thought that the dashboard of the virtual car could be used to show parts of the physical world where it would otherwise provide little to no information in the virtual experience.


In the future, virtual reality experiences could be augmented to react to physical objects. For instance, new physical toys could be designed to act as controllers to the game (e.g., guns, wands, etc.). Designers could also leverage existing objects as weapons, or enable physical interactions with the environment to affect the game. For example, drinking a glass of water can be used to recharge a user’s health in an FPS game.

\balance

\section{Conclusion}

We have explored a design space of bringing physical real world
objects into a virtual reality experience. We selected four renderings
from this design space and compared them through an empirical
evaluation to understand which approaches maximize utility while
reinforcing immersion.  Lastly, we provide critical considerations
necessary for the design of renderings of real world objects in
virtual reality.

\small
\bibliographystyle{acm-sigchi}
\bibliography{paper0.7.7_arxiv}

\begin{thebibliography}{10}

\bibitem{Bai2014}
Bai, H., Lee, G., and Billinghurst, M.
\newblock {Using 3D hand gestures and touch input for wearable AR interaction}.
\newblock {\em Proc. of CHI EA\/} (2014), 1321--1326.

\bibitem{Billinghurst2002}
Billinghurst, M., Kato, H., Kiyokawa, K., Belcher, D., and Poupyrev, I.
\newblock {Experiments with Face-To-Face Collaborative AR Interfaces}.
\newblock {\em Virtual Reality Journal 4}, 2 (2002), 107--121.

\bibitem{Brown2004}
Brown, E., and Cairns, P.
\newblock {A grounded investigation of game immersion}.
\newblock {\em Proc. of CHI\/} (2004), 1297.

\bibitem{Gordon}
Gordon, G., Billinghurst, M., Bell, M., Woodfill, J., Kowalik, B., Erendi, a.,
  and Tilander, J.
\newblock {The use of dense stereo range data in augmented reality}.
\newblock In {\em IEEE ISMAR}, IEEE Comput. Soc (2002), 14--23.

\bibitem{Karl-petter1997}
Karl-petter, K. T.~S.
\newblock {Windows on the World : An example of Augmented Virtuality}.
\newblock In {\em Interfaces} (1997), 482--490.

\bibitem{Kennedy2010494}
Kennedy, R.~S., Drexler, J., and Kennedy, R.~C.
\newblock Research in visually induced motion sickness.
\newblock {\em Applied Ergonomics 41}, 4 (2010), 494 -- 503.
\newblock Special Section - The First International Symposium on Visually
  Induced Motion Sickness, Fatigue, and Photosensitive Epileptic Seizures
  (VIMS2007).

\bibitem{Kreylos2014}
Kreylos, O.
\newblock {Vrui VR Toolkit}, 2014.

\bibitem{Milgram1994}
Milgram, P.
\newblock {Augmented Reality: A class of displays on the reality-virtuality
  continuum}.
\newblock In {\em Proc. of SPIE}, vol.~2351 (1994), 282--292.

\bibitem{Mine1997}
Mine, M.~R., Jr., F. P.~B., and S{\'{e}}quin, C.~H.
\newblock Moving objects in space: Exploiting proprioception in
  virtual-environment interaction.
\newblock In {\em Proceedings of SIGGRAPH 97}, Computer Graphics Proceedings,
  Annual Conference Series (Aug. 1997), 19--26.

\bibitem{Regenbrech2004}
Regenbrech, H.
\newblock {Using Augmented Virtuality for Remote Collaboration}.
\newblock {\em Presence: Teleoperators \& Virtual Environments 13}, 3 (2004),
  338--354.

\bibitem{Schild2012}
Schild, J., LaViola, J., and Masuch, M.
\newblock Understanding user experience in stereoscopic 3d games.
\newblock In {\em Proc. of CHI}, CHI '12, ACM (New York, NY, USA, 2012),
  89--98.

\end{thebibliography}

\end{document}